\documentclass[12pt]{article}

\usepackage{amssymb}
\usepackage{amsmath}

\newcommand{\ben}{\begin{equation}}
\newcommand{\een}{\end{equation}}
\newcommand{\Ai}{\mbox{Ai}}
\renewcommand{\theequation}{\arabic{section}.\arabic{equation}}

\def\Xint#1{\mathchoice
   {\XXint\displaystyle\textstyle{#1}}%
   {\XXint\textstyle\scriptstyle{#1}}%
   {\XXint\scriptstyle\scriptscriptstyle{#1}}%
   {\XXint\scriptscriptstyle\scriptscriptstyle{#1}}%
   \!\int}
\def\XXint#1#2#3{{\setbox0=\hbox{$#1{#2#3}{\int}$}
     \vcenter{\hbox{$#2#3$}}\kern-.5\wd0}}

\def\dashint{\Xint-}

\begin{document}

\begin{titlepage}

\begin{flushright}
RU-NHETC-2010-11
\end{flushright}

\vspace{15mm}

\begin{center}
\LARGE Asymptotic properties of mass spectrum in 't~Hooft's model of mesons
\end{center}

\begin{center}
\large
Iskander Ziyatdinov\footnote{ziyatdin@physics.rutgers.edu}
\vspace{4mm}
\\
NHETC, Department of Physics and Astronomy, 
\\
Rutgers University,
\\ 
Piscataway, NJ 08854--8019, USA
\end{center}

\vspace{10mm}

\begin{abstract}
We study 't Hooft's equation for bound states in 2d multicolor QCD. We consider the case of quarks with equal masses. We derive asymptotic expansions for the spectrum of mesons in different regimes and study their properties.
\end{abstract}

\end{titlepage}

\section{Introduction}

\setcounter{equation}{0}

The 't Hooft model is a name given to the QCD in two dimensions with infinite number of colors \cite{'tHooft:1974hx}. Among many features of the theory we will be interested in the following one: it is shown in \cite{'tHooft:1974hx} that mesons and their spectrum can be exactly described in terms of solutions of a certain integral equation. Although it can be be solved numerically \cite{Hanson:1976ey, Huang:1988br, Jaffe:1991ib, Krauth:1996dg}, this equation, due its elegance and simplicity, deserves an analytical study. One is interested in analytic properties of the spectrum as functions of parameters of theory, that is, the coupling constant and masses of quarks, taken as complex variables. With this goal in mind, the authors  of \cite{Fateev:2009jf} proposed a novel method for studying analytic properties of solutions of the equation, more specifically the spectrum of the problem. Although they consider the case when quark masses have special values, the method they developed is general and many conclusions about properties of the spectrum are applicable or can be extended to a general case without any obvious difficulties.

't Hooft's theory for mesons is not a unique example of theories with similar properties. For example, in \cite{Fonseca:2001dc, Fonseca:2006au} a Bethe-Salpeter equation for the mesons of the Ising field theory (IFT) was proposed. From the particle content point of view, theories are not very different, each being a two-dimensional theory of quarks with a confining interaction. At the same time, whereas the bound state equation in the IFT is written in a certain, rather limiting approximation, the equations that appear in 't Hooft's theory are exact. This fact, together with results of studies of 't Hooft's theory, might give us some intuition as to what to expect in the full IFT.

This work can be seen as a preliminary step in studying the case of quarks with generic masses. Here we restrict ourselves to the case when quarks have equal yet arbitrary masses.  The goal of the paper is to explore asymptotic properties of the spectrum. For this, we develop low-energy and semiclassical expansions for the spectrum, study their properties and compare results with numerical analysis.

This paper is organized as follows. First, we formulate the problem. Next section is devoted to studying the weak-coupling limit, which, depending on certain conditions, leads either to the low-energy or semiclassical expansions. Finally, we compare analytical results with the numerical solution. In the appendix we provide another way to derive the low-energy expansion.


\section{Set-up}

\setcounter{equation}{0}

't Hooft's model, or two-dimensional QCD with gauge group $SU(N)$, is defined by the following Lagrangian 
\ben\label{QCD2}
\mathcal{L}=-\frac{N}{4g^2} \mbox{ Tr }F_{\mu\nu}F^{\mu\nu} +\bar\psi^{(a)}(i\gamma^{\mu}D_\mu-m_0^{(a)})\psi^{(a)},
\een
here $\psi^{(a)}$ are quarks with bare masses $m_0^{(a)}$, different in general; the field strength $F_{\mu\nu}=\partial_\mu A_\nu-\partial_\nu A_\mu+i[A_\mu,\ A_\nu]$ and the covariant derivative $D_\mu=\partial_\mu+iA_\mu$ are defined in terms of the gauge potential $A_\mu$ given by $N\times N$ hermitian traceless matrices. In the large $N$ limit, 't Hooft derived the Bethe-Salpeter equation for mesons, two-quark bound states of the theory\cite{'tHooft:1974hx}. This equation, though originally obtained in the light-cone gauge, is actually gauge-invariant since it describes gauge-invariant objects. When quarks have equal masses $m_0^{(a)}=m_0$, which is the case we will consider, 't Hooft's equation has the form
\ben\label{StartingEquation}
M^2\varphi(x)=\frac{m^2\varphi(x)}{x(1-x)}-\frac{g^2}{\pi}\dashint_0^1\frac{\varphi(y)\ dy}{(y-x)^2}
\een
with $\varphi(x)$ being the wave function of the meson with the mass $M$ and $m^2=m_0^2-g^2/\pi$ standing for the renormalized quark mass. It is this equation, Eq. (\ref{StartingEquation}), that is the main object of our analysis.
Written in terms of dimensionless quantities, $\alpha=M^2/4m^2$ and $\lambda=g^2/2m^2$,  the equation (\ref{StartingEquation}) becomes
\ben\label{xequation}
\alpha\varphi(x)=\frac{\varphi(x)}{4x(1-x)}-\lambda\dashint^1_0\frac{\varphi(y)}{(y-x)^2}\frac{dy}{2\pi}.
\een

The bound-state equation can be viewed as an eigenvalue problem $\hat H\varphi=\alpha\varphi$ for a suitably defined hamiltonian $\hat H$ acting in a certain Hilbert space. Solutions of the problem can be classified according to the symmetry $x\to 1-x$ of the equation.

The overall goal of this work is to provide some insight into analytical properties of the spectrum, that is, properties of $\alpha$ as a function of $\lambda$ taken complex. As we have pointed out in the introduction, at the current stage we will be mainly interested in asymptotic properties of the spectrum leaving other, more interesting questions for later.


\section{Weak-coupling expansion}

\setcounter{equation}{0}

In the weak-coupling limit $\lambda\to0$, each meson's mass $M_n$ will approach $2m$ from above, i.e., $\alpha$ approaches $1_{+0}$. Depending on how $\lambda$ goes to 0, one gets different expansions. For example, if $n$ is kept fixed, this limit will give us the low-energy expansion. Such expansion is known to be in $t=\lambda^{1/3}$. At the same time if one looks at high levels, with $n\geq1/\lambda$, one obtains the semiclassical expansion giving corrections to the already known Bohr--Sommerfeld equation for the spectrum.

In our analysis for consistency we will use notations adopted in \cite{Fonseca:2006au}.

It is convenient to go to the rapidity space,  $\theta=\frac1 2 \log\frac{x}{1-x}$. Then Eq.(\ref{xequation}) takes the form
\ben\label{ThetaEquation}
\left(1-\frac{\alpha}{\cosh^2\theta}\right)\varphi(\theta)=2\lambda\dashint^{\infty}_{-\infty}\frac{d\theta'}{2\pi}\frac{\varphi(\theta')}{\sinh^2(\theta-\theta')}.
\een
Define the hamiltonian $\hat H$ as
\ben
\hat H\varphi(\theta)=\cosh^2\theta\left(\varphi(\theta)-2\lambda\dashint^{\infty}_{-\infty}\frac{d\theta'}{2\pi}\frac{\varphi(\theta')}{\sinh^2(\theta-\theta')}\right),
\een
then Eq.(\ref{ThetaEquation}) becomes equivalent to
\ben
\hat H\varphi(\theta)=\alpha\varphi(\theta)
\een
with $\varphi(\theta)$ belonging to the Hilbert space with the metric
\ben\label{metric}
\|\varphi\|^2=\int^{\infty}_{-\infty}\frac{d\theta}{2\pi}\frac{|\varphi(\theta)|^2}{4\cosh^2\theta}.
\een

For $\varphi(x)$ we use the following ansatz
\ben\label{OddAnsatz}
\varphi^{(0)}_{odd}(\theta)=\int^{\infty}_{-\infty} \frac{\sinh\theta\cosh\beta \ e^{iS(\beta)/\lambda}\ d\beta}{\sinh(\theta+\beta-i0)\sinh(\theta-\beta+i0)}
\een
in the odd sector, and
\ben\label{EvenAnsatz}
\varphi^{(0)}_{even}(\theta)=\int^{\infty}_{-\infty} \frac{(-\sinh\beta\cosh\theta)\ e^{iS(\beta)/\lambda}\ d\beta}{\sinh(\theta+\beta-i0)\sinh(\theta-\beta+i0)}
\een
in the even sector, where
\ben
S(\beta)=\alpha\tanh\beta-\beta.
\een
Both ansatz functions can be shown to be normalizable with respect to the metric (\ref{metric}).

In the further analysis we will be working with the following quantities
\ben
\sinh\theta\ \Delta_{odd}(\theta)\equiv(\hat H-\alpha)\varphi^{(0)}_{odd}(\theta)
\een
and
\ben
-\cosh\theta\ \Delta_{even}(\theta)\equiv(\hat H-\alpha)\varphi^{(0)}_{even}(\theta)
\een

Computation along the lines of the one shown in Appendix B of \cite{Fonseca:2006au} gives the following expressions for these functions
\ben\label{DeltaOdd}
\Delta_{odd}(\theta)=\int^{\infty}_{-\infty} d\beta\ e^{iS(\beta)/\lambda}\left(\frac{\alpha}{\cosh\beta}-\frac{i\lambda\cosh^2\theta\sinh\beta}{(\cosh\beta+\cosh\theta)^2}\right),
\een
\ben\label{DeltaEven}
\Delta_{even}(\theta)=\int^{\infty}_{-\infty} d\beta\ e^{iS(\beta)/\lambda}\left(\frac{\alpha\ \sinh\beta}{\cosh^2\beta}-\frac{i\lambda\cosh\theta\ (\cosh\beta\cosh\theta+1)}{(\cosh\beta+\cosh\theta)^2}\right).
\een

We develop the small-$\lambda$ series for (\ref{DeltaOdd}) and (\ref{DeltaEven}). At first few orders, terms in these expansions will have no dependence on $\theta$, directly giving the spectra $\alpha_n$ as zeroes of $\Delta_{odd/even}(\theta)\equiv\Delta_{odd/even}(\theta|\alpha)$. But in general, these terms will explicitly depend on $\theta$. In such situations one has to proceed as follows. Let $\{\varphi_n\}$ be the complete orthonormal set of eigenvectors of $\hat H$: $\hat H\varphi_n=\alpha_n\varphi_n$. Then the following quantity, set to zero,
\ben\label{SpectralCondition}
C_{n;odd/even}(\alpha)=(\varphi_n,(\hat H-\alpha)\varphi^{(0)}_{odd/even})
\een
will play the role of the spectral condition with $\alpha_n$ being its zeroes. For small values of $\lambda$, $\varphi^{(0)}_{odd/even}$ provides a good approximation for $\varphi_n$. As $\lambda$ grows, corrections can be found by iterating the following expression
\ben
\varphi_n(\theta)=\frac{1}{(\varphi_n,\varphi^{(0)}_{odd/even})}\left(\varphi^{(0)}_{odd/even}(\theta)-\sum_{k\neq n}\frac{C_k(\alpha)}{\alpha_k-\alpha}\varphi_k(\theta)\right).
\een
Finding $\varphi_n$ order by order and plugging it in (\ref{SpectralCondition}), one obtains corrections to the spectral condition. 

\subsection{Low-energy expansion}\label{LowEnergySection}

Assume that $\alpha$ is close to 1 and write it as $\alpha=1+zt^2+\sum_{k=3}e_kt^k$. In this case one sees that the main contribution to the integrals (\ref{DeltaOdd}) and (\ref{DeltaEven})  comes when $\beta\sim t$. Rescaling $\beta=-ut$, $\theta=-vt$, and observing that
\ben
\frac{S(\beta)}{\lambda}=S_0(u)+S_1(u),
\een
where $S_0(u)=\frac{u^3}{3}-zu$, hence giving the Airy function, and $S_1(u)=\frac{S(\beta)}{\lambda}-S_0(u)=O(t)$,
one can expand $\Delta(\theta)$ in $t$. Subsequent setting terms in each order of $t$ equal to zero gives conditions sufficient for determining $\alpha$.

After a straightforward computation one obtains low-energy expansions for $\alpha$ in both sectors. In the odd sector the first terms will have the form
\begin{multline}\label{OddLowEnergyExpansion}
\alpha_{odd}=1+t^2z+\dfrac{t^4z^2}{5}+t^6\left(-\dfrac{3}{175}z^3+\dfrac{6}{35}\right)+
\\
+ t^8\left(\dfrac{23}{7875}z^4-\dfrac{4}{1575}z \right)+\ldots,
\end{multline}
where $z$ is such that $\Ai(-z)=0$. Whereas for even solutions one obtains
\ben\label{EvenLowEnergyExpansion}
\alpha_{even}=1+t^2z+t^4\left(\dfrac{z^2}{5}+\dfrac{1}{5z}\right)+t^6\left(-\dfrac{3}{175}z^3+\dfrac{3}{25}-\dfrac{1}{50z^3}\right)+\ldots
\een
with $z$ being the solution of $\Ai'(-z)=0$.

As above, one should mention that except for the first few terms of the expansion of $\Delta$ in $t$, expressions at powers of $t$ will depend on $v$, that is on $\theta$. In these cases one has to follow the procedure outlined above. We just note that such dependence appears for the first time at $t^8$-order in the odd sector and $t^6$-order in the even.

Another observation is that as opposed to the IFT, $e_{k}$ with odd $k$ seem to be absent in low orders both in the odd and even cases (at least up to $k=15$).

Looking at the coefficients in (\ref{OddLowEnergyExpansion}) and (\ref{EvenLowEnergyExpansion}), one can see the general structure of $e_k$, polynomial in $z$ or in $z$ and $1/z$, respectively. At higher levels $z_n$ behave as $z_n\sim n^{2/3}$ both for odd and even solutions. Hence, the leading terms in $e_k$, terms with the largest power of $z$, which come from the term at $\lambda^0$ in the brackets in (\ref{DeltaOdd}) and (\ref{DeltaEven}), depend on the combination $(n^{1/3}t)^{2m}$. Moreover, one can notice that numerical coefficients are the same for both odd and even sectors. This is expected because the sum of these leading terms is given by the Bohr--Sommerfeld formula, the leading contribution in the semiclassical expansion. Therefore, one would expect that summing sub-leading terms should reproduce contributions of the semiclassical expansion of higher orders.

\subsection{Semiclassics}

When $\lambda$ goes to zero and $\alpha$ is not necessarily close to 1, one obtains the semiclassical approximation. Technically it is reduced to calculating the integrals (\ref{DeltaOdd}) and (\ref{DeltaEven}) using the saddle-point method.

In both sectors the integral for $\Delta(\theta)$ has the form
\ben\label{SaddlePoint}
\Delta(\theta)=\int^{\infty}_{-\infty} d\beta f(\beta|\theta) e^{iS(\beta)/\lambda} .
\een
There are two saddle points  $\beta=\pm\vartheta$ with $\vartheta$ being the positive solution of 
\ben\label{alphan}
\alpha=\cosh^2\vartheta.
\een
With $\Delta_{\pm}(\theta)$ being contributions from $\pm\vartheta$, the integral can be rewritten as
\ben
\Delta(\theta)=\Delta_+(\theta)+\Delta_-(\theta).
\een
In each case, odd or even, one has to compute only $\Delta_+(\theta)$, whereas $\Delta_-(\theta)=\pm\Delta^*_+(\theta)$, here $+/-$ is in the odd and even sectors, respectively.

The leading order calculation yields the following expressions in the odd sector
\ben
\frac{\Delta_{odd}(\theta)\sqrt{\sinh\vartheta}}{2\sqrt{\pi\lambda\cosh^3\vartheta}}=\cos\left(\bar S(\vartheta)/\lambda-\pi/4\right),
\een
and in the even
\ben
\frac{\Delta_{even}(\theta)}{2i\sqrt{\pi\lambda \sinh\vartheta\cosh\vartheta}}=\sin\left(\bar S(\vartheta)/\lambda-\pi/4\right)
\een
with
\ben
\bar S(\vartheta)=\left. S(\beta)\right|_{\beta=\vartheta}=\frac{\sinh2\vartheta-2\vartheta}{2}.
\een
At this level, there is no dependence on $\theta$, and the spectral condition in the leading order will have the Bohr--Sommerfeld form,  $\bar S(\vartheta_n)=\pi\lambda(n-n_0)$, here $n=1$, 2, $\ldots$, $n_0$ is equal either to $1/4$ for the odd sector or $3/4$ for the even.

In contrast to the IFT, $\theta$-dependence becomes relevant already at the one-loop level. Computations at this order give the following expression for the odd case
\ben
\frac{\Delta_{odd}(\theta)\sqrt{\sinh\vartheta}}{2\sqrt{\pi\lambda}\cosh^2\vartheta}=\cos\left(\bar S(\vartheta)/\lambda-\pi/4 +\lambda A_{odd}(\theta|\vartheta)+\lambda S_{1; odd}(\vartheta)\right),
\een
where
\ben
A_{odd}(\theta| \vartheta)=-\frac{\cosh^2\theta\tanh \vartheta}{(\cosh\theta+\cosh\vartheta)^2},
\een
\ben
S_{1; odd}(\vartheta)=\frac{\tanh \vartheta}{4}+\frac{1}{48\tanh^3\vartheta}\left(-9\tanh^4\vartheta+6\tanh^2\vartheta-5\right).
\een
In the even sector the expression is
\begin{multline}
\frac{\Delta_{even}(\theta)}{2i\sqrt{\pi\lambda \sinh\vartheta\cosh\theta}}=
\\
=\sin\left(\bar S(\vartheta)/\lambda-\pi/4+\lambda A_{even}(\theta|\vartheta)+\lambda S_{1; even}(\vartheta)\right)
\end{multline}
with
\ben
A_{even}(\theta|\vartheta)=-\frac{\cosh\theta(\cosh\theta\cosh\vartheta+1)}{\sinh\vartheta(\cosh\theta+\cosh\vartheta)^2},
\een
\ben
S_{1; even}(\vartheta)=\frac{1}{4\tanh^3\vartheta}+\frac{1}{48\tanh^3\vartheta}\left(-9\tanh^4\vartheta+6\tanh^2\vartheta-5\right).
\een

In this situation one again has to use the method based on (\ref{SpectralCondition}). At this order, when $\varphi_n$ is approximated by $\varphi^{(0)}_{odd/even}$, the integral in (\ref{SpectralCondition}) at small $\lambda$ will have the main contribution coming from the saddle points $\theta=\pm\vartheta$ giving the following quantization conditions in the odd sector
\ben\label{OddSemiclassics}
\frac{\sinh2\vartheta_n-2\vartheta_n}{2}=\pi\lambda(n-1/4)-\lambda^2\left(A_{odd}(\vartheta_n|\vartheta_n)+S_{1; odd}(\vartheta_n)\right)+O(\lambda^3)
\een
and in the even sector
\begin{multline}\label{EvenSemiclassics}
\frac{\sinh2\vartheta_n-2\vartheta_n}{2}=\pi\lambda(n-3/4)-
\\
-\lambda^2\left(A_{even}(\vartheta_n|\vartheta_n)+S_{1; even}(\vartheta_n)\right)+O(\lambda^3).
\end{multline}
Once $\vartheta_n$ is found, the eigenvalues $\alpha_n$ are determined using (\ref{alphan}).

Calculation in higher orders is technically similar, albeit more involved.

In principle, using (\ref{OddSemiclassics}) and (\ref{EvenSemiclassics}) one can re-derive low-energy expansions (\ref{OddLowEnergyExpansion}) and (\ref{EvenLowEnergyExpansion}), respectively.


\section{Numerical results}

\setcounter{equation}{0}

Over the years there have been proposed different approaches for numerical solution of the problem, each giving solution to 't Hooft's equation with any degree of accuracy. In \cite{Fonseca:2006au} another method, based on discretizing the Fourier-transformed version of 't Hooft's equation, was suggested. Applying the Fourier transform to Eq.(\ref{ThetaEquation})
\ben
\varphi(\theta)=\int^{\infty}_{-\infty} d\nu \ e^{-i\nu\theta}\varphi(\nu), \quad \varphi(\nu)=\int^{\infty}_{-\infty} \frac{d\theta}{2\pi}\ e^{i\nu\theta}\varphi(\theta),
\een
one obtains
\ben\label{FourierRapidity}
\left(1+\lambda\nu\coth\frac{\pi\nu}{2}\right)\varphi(\nu)=\alpha\int^{\infty}_{-\infty} d\nu'\ K(\nu-\nu')\varphi(\nu')
\een
with $K(\nu)=\nu/(2\sinh\frac{\pi\nu}{2})$. The main feature of this version of 't Hooft's equation is the smoothness of its kernel allowing to discretize the equation directly. Even though this method can be not as efficient as others, its ease makes up for it. One should mention that this is not the first time this parametrization has been used. It has already appeared, for example, in \cite{Narayanan:2005gh} and recently in \cite{Fateev:2009jf}.

In Tables 1 and 2 we list values for $\alpha$, i.e., for $M^2$ in units of $4m^2$, for first few levels, both even and odd, at different values of $\lambda$. Numerical results are obtained by discretizing Eq.(\ref{FourierRapidity}) with the number of steps $N=3000$ and step size $\varepsilon=0.1$. This choice of discretization parameters provides more than sufficient accuracy in this range of $\lambda$, at least for the first levels. They are compared to the results obtained by means of derived expansions (\ref{OddLowEnergyExpansion}) and (\ref{EvenLowEnergyExpansion}), truncated at $\lambda^{10/3}$ in the odd sector and at $\lambda^{8/3}$ in the even sector. One can see that within the given accuracy numerical results match the expected values with deviations determined by the order of the terms dropped from the low-energy expansion.
\begin{table}[h]\label{OddResults}
\centering
\small
\caption{$\alpha_n$ for first odd levels. The first value is numerical; the second is obtained using the truncated low-energy expansion (\ref{OddLowEnergyExpansion}); the third value is the order of the difference between the two.}
\vspace{2mm}
\begin{tabular}{c|c|c|c|c}
$\lambda$ &    0.0001 & 0.001 & 0.01 & 0.1 \\
\hline
\hline
$n=1$ &
\begin{tabular}{c}
1.00504237411651\\
1.00504237411666\\
$10^{-13}$
\end{tabular}
&
\begin{tabular}{c}
 1.02349036201\\
 1.02349036216 \\
 $10^{-10}$
\end{tabular}
 &
 \begin{tabular}{c}
 1.11087634\\
 1.11087649\\
 $10^{-7}$
\end{tabular}
 &
 \begin{tabular}{c}
 1.55404\\
 1.55418 \\
 $10^{-4}$
\end{tabular}
 \\
\hline
\hline
$n=2$ &
 \begin{tabular}{c}
 1.00882272353899\\
 1.00882272353917 \\
 $10^{-13}$
\end{tabular}
&
 \begin{tabular}{c}
 1.04121272921\\
 1.04121272942 \\
 $10^{-10}$
\end{tabular}
&
 \begin{tabular}{c}
 1.19684999\\
 1.19685027 \\
 $10^{-7}$
\end{tabular}
&
 \begin{tabular}{c}
 2.02721\\
 2.02759 \\
 $10^{-4}$
\end{tabular}
\\
\hline
\hline
$n=3$ &
 \begin{tabular}{c}
 1.01192195048315\\
 1.01192195048344 \\
 $10^{-13}$
\end{tabular}
&
 \begin{tabular}{c}
 1.05581244361\\
 1.05581244407 \\
 $10^{-10}$
\end{tabular}
&
 \begin{tabular}{c}
 1.26911411\\
 1.26911489 \\
 $10^{-6}$
\end{tabular}
&
 \begin{tabular}{c}
 2.44971\\
 2.45097 \\
 $10^{-4}$
\end{tabular}
\\
\hline
\hline
$n=4$
&
 \begin{tabular}{c}
 1.01466422536562\\
 1.01466422536618 \\
 $10^{-12}$
\end{tabular}
&
 \begin{tabular}{c}
 1.06878314251\\
 1.06878314354 \\
 $10^{-9}$
\end{tabular}
&
 \begin{tabular}{c}
 1.33436550\\
 1.33436743 \\
 $10^{-6}$
\end{tabular}
&
 \begin{tabular}{c}
 2.84797\\
 2.85117 \\
 $10^{-3}$
\end{tabular}
\\
\hline
\hline
$n=5$
&
 \begin{tabular}{c}
 1.01717361844767 \\
 1.01717361844723 \\
 $10^{-13}$
\end{tabular}
&
 \begin{tabular}{c}
 1.08069521197 \\
 1.08069521405 \\
 $10^{-9}$
\end{tabular}
&
 \begin{tabular}{c}
 1.39513452 \\
 1.39513856 \\
 $10^{-6}$
\end{tabular}
&
 \begin{tabular}{c}
 3.23152 \\
 3.23815 \\
 $10^{-2}$
\end{tabular}
\end{tabular}
\end{table}
\begin{table}[h]\label{EvenResults}
\centering
\small
\caption{Same as in Table 1, but for first even levels.}
\vspace{2mm}
\begin{tabular}{c|c|c|c|c}
$\lambda$  & 0.0001 & 0.001 & 0.01 & 0.1 \\
\hline
\hline
$n=1$
&
 \begin{tabular}{c}
 1.002196798442 \\
 1.002196798481 \\
 $10^{-11}$
\end{tabular}
&
 \begin{tabular}{c}
 1.0102283943 \\
 1.0102284025 \\
 $10^{-8}$
\end{tabular}
&
 \begin{tabular}{c}
 1.0481649 \\
 1.0481666 \\
 $10^{-6}$
\end{tabular}
&
 \begin{tabular}{c}
 1.2387 \\
 1.2391 \\
 $10^{-4}$
\end{tabular}
\\
\hline
\hline
$n=2$
&
 \begin{tabular}{c}
 1.007008105144 \\
 1.007008105149 \\
 $10^{-12}$
\end{tabular}
&
 \begin{tabular}{c}
 1.0326986818 \\
 1.0326986807 \\
 $10^{-9}$
\end{tabular}
&
 \begin{tabular}{c}
 1.1554011 \\
 1.1554000 \\
 $10^{-6}$
\end{tabular}
&
 \begin{tabular}{c}
 1.7964 \\
 1.7959 \\
 $10^{-3}$
\end{tabular}
\\
\hline
\hline
$n=3$
&
 \begin{tabular}{c}
 1.010406331508 \\
 1.010406331481 \\
 $10^{-11}$
\end{tabular}
&
 \begin{tabular}{c}
 1.0486680231 \\
 1.0486680086 \\
 $10^{-8}$
\end{tabular}
&
 \begin{tabular}{c}
 1.2336564 \\
 1.2336495 \\
 $10^{-5}$
\end{tabular}
&
 \begin{tabular}{c}
 2.2408 \\
 2.2381 \\
 $10^{-3}$
\end{tabular}
\\
\hline
\hline
$n=4$
&
 \begin{tabular}{c}
 1.013313818353 \\
 1.013313818267 \\
 $10^{-10}$
\end{tabular}
&
 \begin{tabular}{c}
 1.0623921932 \\
 1.0623921521 \\
 $10^{-8}$
\end{tabular}
&
 \begin{tabular}{c}
 1.3021422 \\
 1.3021238 \\
 $10^{-5}$
\end{tabular}
&
 \begin{tabular}{c}
 2.6502 \\
 2.6430 \\
 $10^{-2}$
\end{tabular}
\\
\hline
\hline
$n=5$
&
 \begin{tabular}{c}
 1.015933386177 \\
 1.015933385994 \\
 $10^{-10}$
\end{tabular}
&
 \begin{tabular}{c}
 1.0748048015 \\
 1.0748047167 \\
 $10^{-7}$
\end{tabular}
&
 \begin{tabular}{c}
 1.3650255 \\
 1.3649880 \\
 $10^{-5}$
\end{tabular}
&
 \begin{tabular}{c}
 3.0406 \\
 3.0266 \\
 $10^{-2}$
\end{tabular}
\end{tabular}
\end{table}

It is worthwhile to see how the expansions (\ref{OddLowEnergyExpansion}) and (\ref{EvenLowEnergyExpansion})  approach the numerical values. For the lowest odd and even levels, which we choose because zeroes of $\Ai(-z)$ and $\Ai'(-z)$ are smallest for them thus making specific features of the expansions, especially in the even case, more prominent, this can be observed in Tables 3 and 4, respectively. Here $\alpha^{(k)}_{odd/even}$ denotes estimates obtained using (\ref{OddLowEnergyExpansion}) and (\ref{EvenLowEnergyExpansion}) by keeping terms up to $t^k$ included.
\begin{table}[h]\label{OddApprox}
\small
\centering
\caption{Step-by-step approximation of the meson mass for the the lowest odd level obtained by successively including higher terms in (\ref{OddLowEnergyExpansion}); the last value is the numerical result.}
\vspace{2mm}
\begin{tabular}{c|c|c|c|c}
$\lambda$  & 0.0001 & 0.001 & 0.01 & 0.1 \\
\hline
\hline
$\alpha^{(2)}_{odd}$    & 1.00503729971412 &  1.02338107410 &  1.10852533  & 1.50373 \\
$\alpha^{(4)}_{odd}$    & 1.00504237459180 &  1.02349040903 &  1.11088088  & 1.55448 \\
$\alpha^{(6)}_{odd}$    & 1.00504237411491 &  1.02349036134 &  1.11087611  & 1.55400 \\
$\alpha^{(8)}_{odd}$    & 1.00504237411666 &  1.02349036215 &  1.11087649  & 1.55418 \\
\hline
$\alpha^{(num)}_{odd}$&1.00504237411651 &  1.02349036201 &  1.11087634  & 1.55404 \\
\end{tabular}
\end{table}
\begin{table}[h]\label{EvenApprox}
\small
\centering
\caption{Same as in Table 3, but for the lowest even level.}
\vspace{2mm}
\begin{tabular}{c|c|c|c|c}
$\lambda$  & 0.0001 & 0.001 & 0.01 & 0.1 \\
\hline
\hline
$\alpha^{(2)}_{even}$    & 1.002194922920 &  1.010187930 &  1.0472882 & 1.21949 \\
$\alpha^{(4)}_{even}$    & 1.002196797651  & 1.010228320 &  1.0481584 & 1.23824 \\
$\alpha^{(6)}_{even}$    & 1.002196798481  & 1.010228403 &  1.0481666 & 1.23907 \\
\hline
 $\alpha^{(num)}_{even}$ &1.002196798442  & 1.010228394 &  1.0481649 & 1.23870 \\
\end{tabular}
\end{table}

Tables 5 and 6 contain values of $\alpha_n$ obtained through semiclassical expansions, (\ref{OddSemiclassics}) and (\ref{EvenSemiclassics}). As expected, with $n$ increasing, semiclassical results approach values obtained through discretization.
\begin{table}[h]\label{OddDataSemiclassics}
\small
\centering
\caption{$\alpha_n$ for first odd levels from semiclassics.}
\vspace{2mm}
\begin{tabular}{c|c|c|c|c}
$\lambda$  & 0.0001 & 0.001 & 0.01 & 0.1 \\
\hline
\hline
$n=1$ & 1.005045002 & 1.023502645  & 1.110935189 & 1.554453706 \\
$n=2$ & 1.008822931 & 1.041213704  & 1.196854870 & 2.027350588 \\
$n=3$ & 1.011921999 & 1.055812673  & 1.269115381 & 2.449829823 \\
$n=4$ & 1.014664243 & 1.068783226  & 1.334366056 & 2.848080751 \\
$n=5$ & 1.017173627 & 1.080695251  & 1.395134852 & 3.231630623
\end{tabular}
\end{table}
\begin{table}[h]\label{EvenDataSemiclassics}
\small
\centering
\caption{$\alpha_n$ for first even levels from semiclassics.}
\vspace{2mm}
\begin{tabular}{c|c|c|c|c}
$\lambda$  & 0.0001 & 0.001 & 0.01 & 0.1 \\
\hline
\hline
$n=1$ & 1.002079962 & 1.009690485 & 1.045761963 & 1.229713753 \\
$n=2$ & 1.007007280 & 1.032694849 & 1.155383793 & 1.796538843 \\
$n=3$ & 1.010406206 & 1.048667441 & 1.233654091 & 2.240958482 \\
$n=4$ & 1.013313781 & 1.062392019 & 1.302141769 & 2.650298641 \\
$n=5$ & 1.015933371 & 1.074804730 & 1.365025460 & 3.040735165
\end{tabular}
\end{table}


\section{Conclusion}

In this work we apply methods proposed in \cite{Fonseca:2001dc, Fonseca:2006au} to the study of 't Hooft's model. In particular, we study asymptotic properties of the spectrum of mesons and obtain the low-energy and semiclassical expansions for meson masses. We compare these results to the numerical solution of the problem. 

This study has to be considered as a part of an extended project. Methods used in this work, together with the approach proposed in \cite{Fateev:2009jf}, will likely provide a way to tackle the general case of the problem with arbitrary quark masses. The project itself has many possible ramifications. Some of them are related to questions about analytical properties of the spectra which are preserved if one considers the case of number of colors large but finite. In such cases one is primarily interested in the nature of singularities. Others deal with possible generalizations of these methods to more general field theories. For example, one can consider theories with vacuum degeneracy, possessing certain integrabilty which becomes broken by perturbations that typically give rise to the confining force between particles of original theories. IFT and 't Hooft's theory provide sufficiently simple examples of such theories.


\section*{Acknowledgements}

\noindent I would like to thank A.~Zamolodchikov for useful discussions. This work is supported in part by DOE grant DE-FG02-96ER40949.


\appendix

\section{Coordinate-space representation}

\renewcommand{\theequation}{\Alph{section}.\arabic{equation}}

\setcounter{equation}{0}

Dependence of the terms in the low-energy expansion on roots of $\Ai'(-z)$ in (\ref{EvenLowEnergyExpansion}) looks drastically different from the behavior in the odd sector. Therefore, it is worthwhile to provide another way to reproduce it. 

In \cite{Fonseca:2001dc} a coordinate-space formulation for the bound-state equation was used to obtain the low-energy series for the spectrum of the IFT. Similarly we map 't Hooft's equation into the coordinate space. For simplicity we choose Eq.(\ref{xequation}) as a starting point for our analysis. In $u=2x-1$, this equation becomes
\ben
\alpha\varphi(u)=\frac{\varphi(u)}{1-u^2} - 2\lambda\dashint_{-1}^{1}\frac{\varphi(v)}{(u-v)^2}\frac{dv}{2\pi}.
\een
Rescaling $u=tk$, $v=tp$, setting $\lambda=t^3$, expanding everything in $t$, and rearranging the terms, one gets
\ben
(1-\alpha+t^2k^2+t^4k^4+\dots)\varphi(tk) = 2t^2\dashint_{-\infty}^{\infty}\frac{\varphi(tp)}{(p-k)^2}\frac{dp}{2\pi}.
\een
Here we changed the limits of integration from $(-1/t, 1/t)$ to $(-\infty,\infty)$, an operation legitimate at $O(t^4)$. 

Now this equation in the configuration space becomes
\ben\label{SimpleVersion}
\left(-\varepsilon+|x|-\frac{d^2}{dx^2}+t^2\frac{d^4}{dx^4}-t^4\frac{d^6}{dx^6}+\dots\right)\psi(x)=0.
\een
Here the Fourier transform conventions in terms of the rescaled distance between quarks $x=tx_{real}$ are
$$
\psi(x)=\int^{\infty}_{-\infty}\frac{dk}{2\pi} e^{ikx}\varphi(tk),
$$
and we let $1-\alpha=-\varepsilon t^2$.

This equation (\ref{SimpleVersion}) has a symmetry $x\to-x$, hence its spectrum can divided into even and odd parts.

To find a perturbative solution for it, one first solves the auxiliary equation 
\ben\label{Fequation}
\left(x-\frac{d^2}{dx^2}+t^2\frac{d^4}{dx^4}-t^4\frac{d^6}{dx^6}+\dots\right)F(x)=0
\een
with 
$$
F(x)=F_0(x)+t^2F_1(x)+t^4F_2(x)+\dots
$$
The bounded perturbative solution for (\ref{Fequation}) is
\begin{multline}\label{F}
F(x)=\Ai(x)+\frac{t^2}{5}\left(4x\, \Ai(x)+x^2\, \Ai'(x)\right)+
\\
+t^4\left(\left(\frac{x^5}{50}+\frac{5x^2}{7}\right)\Ai(x)+\left(\frac{9x^3}{35}+\frac{6}{35}\right)\Ai'(x)\right)+\ldots
\end{multline}

The following two claims provide both the solutions for (\ref{SimpleVersion}) and the spectra for even and odd sectors. In each case we assume that $F(x)$ is the perturbative solution (\ref{F}) for (\ref{Fequation}).

\textbf{Claim 1.} $\psi_{odd}(x)=\mbox{sgn}\, x\ F(|x|-\varepsilon)$ provides the odd perturbative solution of (\ref{SimpleVersion}) if $F(x)$ is such that
\ben\label{ConditionsOdd}
\begin{array}{l}
F(-\varepsilon)=O(t^2),
\\
F(-\varepsilon)-t^2F''(-\varepsilon)=O(t^4),
\\
F(-\varepsilon)-t^2F''(-\varepsilon)+t^4F^{\rm IV}(-\varepsilon)=O(t^6).
\end{array}
\een

\textbf{Claim 2.} $\psi_{even}(x)=F(|x|-\varepsilon)$ is the even perturbative solution of (\ref{SimpleVersion}) if $F(x)$ satisfies
\ben\label{ConditionsEven}
\begin{array}{l}
F'(-\varepsilon)=O(t^2),
\\
F'(-\varepsilon)-t^2F'''(-\varepsilon)=O(t^4),
\\
F'(-\varepsilon)-t^2F'''(-\varepsilon)+t^4F^{ \rm V}(-\varepsilon)=O(t^6).
\end{array}
\een

The proof of each claim is straightforward. Higher order conditions can be easily written but they are not necessary at the order we are interested in. 

In both cases conditions on $F$, (\ref{ConditionsOdd}) and (\ref{ConditionsEven}), are sufficient for determining the spectra. It is easy to check that using them one obtains  the following expansion of $\alpha$ in the odd sector
\ben
\alpha_{odd}=1+t^2z+\frac{t^4z^2}{5}+t^6\left(-\frac{3}{175}z^3+\frac{6}{35}\right)+\ldots
\een
and in the even sector
\ben
\alpha_{even}=1+t^2z+t^4\left(\frac{z^2}{5}+\frac{1}{5z}\right)+t^6\left(-\frac{3}{175}z^3+\frac{3}{25}-\frac{1}{50z^3}\right)+\ldots
\een
where $z$ is the solution of  $\Ai(-z)=0$ and  $\Ai'(-z)=0$, respectively, thus matching low-energy expansions (\ref{OddLowEnergyExpansion}) and (\ref{EvenLowEnergyExpansion}) derived above at the chosen order in $t$.


\end{document}